\documentclass[10pt,a4paper]{book}

\usepackage{frontier07}

\newcommand{\cf}{\textit{cf.}}

\newcommand{\beq}{\begin{equation}}
\newcommand{\eeq}{\end{equation}}

\newcommand{\ben}{\begin{eqnarray}}
\newcommand{\een}{\end{eqnarray}}

\newcommand{\dvecx}{{\ifmmode d^3 \vec{x} \else $d^3 \vec{x}$\fi}}
\newcommand{\vecx}{{\ifmmode \vec{x} \else $\vec{x}$\fi}}

\newcommand{\rhosm}{{\ifmmode \rho_{\rm sm} \else $\rho_{\rm sm}$\fi}}

\newcommand{\mchi}{\ifmmode {m_{\chi}} \else $m_{\chi}$\fi}

\newcommand{\pos}{{\ifmmode e^+ \else $e^+$\fi}}
\newcommand{\poss}{{\ifmmode {e^{+}}{\rm 's} \else ${e^{+}}{\rm 's}$\fi}}
\newcommand{\pbar}{{\ifmmode \overline{p} \else $\overline{p}$\fi}}
\newcommand{\pbars}{{\ifmmode \overline{p}{\rm 's} \else 
    $\overline{p}{\rm 's}$\fi}}

\newcommand{\gtilde}{{\ifmmode \tilde{\cal G} \else $\tilde{{\cal G}}$\fi}}

\newcommand{\Msol}{{\ifmmode M_{\odot} \else $M_{\odot}$\fi}}
\newcommand{\Rsol}{{\ifmmode R_{\odot} \else $R_{\odot}$\fi}}
\newcommand{\xsol}{{\ifmmode \vec{x}_{\odot} \else $\vec{x}_{\odot}$\fi}}
\newcommand{\rhosol}{{\ifmmode \rho_{\odot} \else $\rho_{\odot}$\fi}}
\newcommand{\fsol}{{\ifmmode f_{\odot} \else $f_{\odot}$\fi}}

\newcommand{\Dvir}{{\ifmmode \Delta_{\rm vir} \else $\Delta_{\rm vir}$\fi}}
\newcommand{\cvir}{{\ifmmode c_{\rm vir} \else $c_{\rm vir}$\fi}}
\newcommand{\Rvir}{{\ifmmode R_{\rm vir} \else $R_{\rm vir}$\fi}}
\newcommand{\Rvirh}{{\ifmmode R_{\rm vir}^{\rm h} \else 
    $R_{\rm vir}^{\rm h}$\fi}}
\newcommand{\Mmin}{{\ifmmode M_{\rm min} \else $M_{\rm min}$\fi}}
\newcommand{\Mmax}{{\ifmmode M_{\rm max} \else $M_{\rm max}$\fi}}

\newcommand{\Mvir}{{\ifmmode M_{\rm vir} \else $M_{\rm vir}$\fi}}
\newcommand{\Mvirh}{{\ifmmode M_{\rm vir}^{\rm h} \else 
    $M_{\rm vir}^{\rm h}$\fi}}
\newcommand{\alpham}{{\ifmmode \alpha_{\rm m} \else $\alpha_{\rm m}$\fi}}

\newcommand{\rhocrit}{{\ifmmode \rho_{\rm crit} \else $\rho_{\rm crit}$\fi}}
\newcommand{\rhocl}{{\ifmmode \rho_{\rm cl} \else $\rho_{\rm cl}$\fi}}
\newcommand{\Mcl}{{\ifmmode M_{\rm cl} \else $M_{\rm cl}$\fi}}
\newcommand{\Mcltot}{{\ifmmode M_{\rm cl}^{\rm tot} \else 
    $M_{\rm cl}^{\rm tot}$\fi}}
\newcommand{\Ncl}{{\ifmmode N_{\rm cl} \else $N_{\rm cl}$\fi}}
\newcommand{\ncl}{{\ifmmode n_{\rm cl} \else $n_{\rm cl}$\fi}}
\newcommand{\phicl}{{\ifmmode \phi_{\rm cl} \else $\phi_{\rm cl}$\fi}}
\newcommand{\phicltot}{{\ifmmode \phi_{\rm cl}^{\rm tot} \else 
    $\phi_{\rm cl}^{\rm tot}$\fi}}
\newcommand{\Beff}{{\ifmmode B_{\rm eff} \else $B_{\rm eff}$\fi}}

\newcommand{\prob}{{\ifmmode {\cal P} \else ${\cal P}$\fi}}

\newcommand{\kpc}{{\ifmmode {\rm kpc} \else ${\rm kpc}$\fi}}

\begin{document}
\newcounter{ctr}
\setcounter{ctr}{\thepage}
\addtocounter{ctr}{8}

\talktitle{Indirect detection of Dark Matter with antimatter: Demystifying 
  the clumpiness boost factors}

\talkauthors{
  J.~Lavalle\footnote{lavalle@in2p3.fr OR lavalle@to.infn.it} 
  \structure{a}\structure{b}
}

\begin{center}
\authorstucture[a]{
  Centre de Physique des Particules de Marseille -- CPPM,\newline
  CNRS-IN2P3 / Universit\'e de la M\'editerran\'ee,\newline
  163, Avenue de Luminy -- mailbox 902,
  F-13288 Marseille cedex 09 -- France.
}
\authorstucture[b]{
  Dipartimento di Fisica Teorica,
  Universit\`a di Torino,\newline
  Via Giuria 1 -- mailbox 150,
  10125 Torino -- Italia.
}

\end{center}

\shorttitle{Demystifying the clumpiness boost factors} 
\firstauthor{J.~Lavalle}

\vspace{-2.5cm}
\begin{abstract}
The hierarchical scenario of structure formation, in the frame of the
$\Lambda$-CDM cosmology, predicts the existence of dark matter (DM) sub-halos 
down to very small scales, of which the minimal size depends on the 
microscopic properties of the DM. In the context of annihilating DM, such 
substructures are expected to enhance the primary cosmic ray (CR) fluxes 
originating from DM annihilation in the Galaxy. This enhancement has long 
been invoked to allow predictions of imprints of DM annihilation on the 
antimatter CR spectra. Taking advantage of the method developed 
by~\cite{boost_method_lavalle_etal_07}, 
we~\cite{boost_clumps_lavalle_etal_07} accurately compute the boost factors 
for positrons and anti-protons, as well as the associated theoretical and 
statistical errors. To this aim, we use a compilation of the latest 
results of cosmological N-body simulations and the theoretical insights found 
in the literature. We find that sub-halos are not likely to significantly 
boost the exotic production of antimatter CRs.
\end{abstract}
\paragraph{Introduction:}
Like other topics in fundamental physics, 
annihilating dark matter is rather well motivated by some \emph{coincidence} 
arguments, because microscopic new physics is expected to surge at the 
electroweak energy scale: This energy scale gives also the right 
interaction amplitudes to naturally get the correct abundance of WIMPs today 
to account for the cosmological DM energy density, provided that they are 
thermally produced in the early Universe, without 
matter-antimatter asymmetry (for a recent review on dark matter candidates, 
see the nice lecture by~\cite{review_dm_murayama_07}). The dark matter 
annihilation properties are intimately connected to its 
possible direct interaction with normal matter, and this sketches two 
complementary approaches in order to hunt for dark matter signatures, beside 
its production in particle colliders. Direct detection experiments focus on 
energy deposits on very sensitive and deep underground detectors, while 
indirect detection aims at observing annihilation traces in the form of 
neutral or charged cosmic rays (for a review on indirect detection, 
see e.g.~\cite{review_dm_carr_etal_06}). Primary antimatter CRs originating 
from DM annihilation could be hunted e.g. in the positron, anti-proton and 
anti-deuteron spectra at the Earth~\cite{salati_conf_06}. In particular, the 
positron spectrum has some features around 10 GeV that are still hardly 
understandable from a secondary origin~\cite{beatty_etal_04}. Some authors 
tried to explain this with DM annihilation induced positrons, 
but most of the predictions rely on a rescaling of the primary fluxes by 
invoking some constant \emph{boost factors} due to DM clumpiness 
(see e.g.~\cite{baltz_edsjo_99,2006JCAP...12..003M}). 
Indeed, many DM substructures are predicted in CDM-type cosmologies, and as 
the annihilation rate is proportional to the squared DM density, it 
naturally increases in an inhomogeneous medium, for a given average 
density. Though the idea of enhancement due to sub-halos is far from 
new~\cite{1993ApJ...411..439S}, it has never been fully addressed 
in the context of charged cosmic rays, contrary to gamma-rays, due to the 
additional difficulty of dealing with CR propagation. Furthermore, because 
dark matter clumps might be numerous and countable (at least 
theoretically or in simulations), they are better described by some phase 
space distributions. The full treatment of this problem has therefore to 
make use of statistics, and should also provide either predictions and the 
associated statistical uncertainties (beside the theoretical ones). 
As a matter of fact, from an observational point of view, any signature can be 
said unambiguous if, at least, the theoretical predictions do not suffer from 
large uncertainties, unless the expected signature is very specific and clear. 
\cite{boost_method_lavalle_etal_07} developed a detailed method in order to 
translate the clumpiness phase space in terms of probability for the primary 
signal/boost. These authors focused on positrons, and took a very simple model 
in which all sub-halos had the same properties and were spatially distributed 
according to the host DM halo profile. They showed that the boost factor and 
the statistical uncertainties actually non trivially depends on the detected 
CR energy. We have used their method, which is quite general and applies to 
any CR species, but with more sophisticated 
and more precise inputs for the dark matter distribution. This proceeding 
summarises the work presented in~\cite{boost_clumps_lavalle_etal_07}, to which 
we refer the reader for a more exhaustive bibliography, and in which we 
accurately compute the boost factors resulting from dark matter clumpiness for 
positrons (\poss) and anti-protons (\pbars).
\paragraph{Dark Matter distribution in the Galaxy:}
Any prediction for indirect detection of DM strongly depends on the 
DM distribution that is used, for the smooth halo as well as for sub-halos (we 
refer to the \emph{smooth} dark matter component as the host DM 
halo of the Milky Way | MW). The density profile of dark objects is the 
main quantity to anticipate how the DM annihilation rate may spatially evolve 
in the Galaxy. Though axysymmetry characterises most of the results of 
cosmological N-body simulations, we adopt here a spherical symmetry to 
describe the Galactic host halo and the sub-halos. A structure of any scale 
may thus have its density profile $\rho(r)$ depicted by the generic formula 
proposed by~\cite{zhao_96}:
$\rho (r) = \rho_s/\left[\left(r/r_s\right)^{\gamma}
  \lbrack 1+(r/r_s)^{\alpha} \rbrack^{(\beta-\gamma)/\alpha} \right]$,
where $\rho_s$ is twice the density at the scale radius $r_s$, at which the 
logarithmic slope changes from $\gamma$ to $(\beta-\gamma)/\alpha$. The 
well-known NFW~\cite{nfw_97} and Moore~\cite{moore_98} profiles are recovered 
with $\{\alpha,\beta,\gamma\} = \{1,3,1 \}$ and $\{1.5,3,1.5 \}$ respectively.
An NFW behaves like $r^{-1}$ while a 
Moore scales like $r^{-1.5}$ in the central region of the structure. The 
scale radius is connected to the \emph{concentration} parameter that we will 
discuss further. The DM density may saturate at the very centre of a structure 
because of annihilation, and a cut-off radius is usually set by equating the 
gravitational infall rate with the WIMP annihilation rate, of which the 
typical size is $\sim 10^{-6}$ pc for usual WIMPs. For the smooth component, 
we choose $r_s = 20$ kpc, and we normalise the density profile \rhosm\ with 
respect to the local DM density in the solar neighbourhood 
$\rhosm(\Rsol = 8\;{\rm kpc}) = \rhosol = 0.3$ GeV/cm$^3$. The inner shape of 
the smooth component has no significant impact on the DM annihilation induced 
CRs because of propagation effects that strongly dilute what originates from 
the central regions of the Galaxy. Fluxes at the Earth are much more sensitive 
to modifications of local distribution of DM. Finally, we set the Galactic 
virial radius to \Rvirh = 280 kpc, so that the total halo mass inside \Rvirh\ 
is $\Mvirh = 1.12\times 10^{12}\Msol$. Sub-halos are found to be more and more 
numerous as the resolution of 
N-body simulations gets thinner and thinner. For instance~\cite{diemand_05} 
found $\sim 10^{15}$ substructures of Earth-mass and of solar system size, 
in a galaxy-size box at early stages of structure formation ($z=26$). Such 
small masses are consequent of very small free streaming scales for the 
dark matter, and are rather generic for WIMPs. However, one can wonder whether 
or not such small structures may survive tidal effects and encounter events 
that characterise their evolution in the host halo. While this issue is 
still debated, we have scrutinised the effect of considering different 
minimal masses | $10^{-6}$, 1 and $10^{6}$ \Msol | in the original paper. 
A complete sub-halo phase space may be portrayed with the following normalised 
probability density functions (pdf), 
which define the number density of sub-halos per mass unit in the MW:
$\frac{d\ncl}{dM}(\vecx,M) = \frac{d\Ncl}{dV\; dM} = N_0 \times 
\frac{d\prob_V (\vecx)}{dV} \times \frac{d\prob_M (M)}{dM}\;,$
where we assume that the mass function has no spatial dependence. The mass 
function is usually found to be a power law ; we therefore define:
$\frac{d\prob_M(M)}{dM} \equiv K_M M^{-\alpham}\;,$
where $K_M$ normalises the pdf to unity in the mass range used for sub-halos, 
so that it depends on \Mmin, \Mmax\ and \alpham. For simplicity, we have 
considered the same power law all over the mass range, which is not that 
obvious but motivated by theoretical 
arguments~\cite{review_mass_function_popolo_07} and simulation 
results~\cite{diemand_05}. Regarding the spatial pdf, we have retained two 
cases. The first one is that sub-halos spatially stick to the 
smooth host DM halo density profile, which may be an appropriate description 
for very light clumps, as they may behave like test particles. In that case, 
$d\prob_V(\vecx)/dV = \rhosm(\vecx)/\Mvirh$, which is normalised inside 
\Rvirh. Nevertheless, according to the current results of N-body simulations 
for which the resolved sub-halo mass is $\gtrsim 10^6 \Msol$ (\cf~e.g. the 
\emph{Via Lactea} \cite{diemand_via_lactea}), the spatial distribution of 
substructures is mostly anti-biased compared to the smooth component. This is 
characterised by a spherical isothermal distribution with a core radius $r_c$ 
roughly equal to the scale radius $r_s$ of the MW. This is the 
2$^{\rm nd}$ case:
$\frac{d\prob_V (r)}{dV} = K_V \times \left[ 1+
  \left(\frac{r}{r_c}\right)^2\right]^{-1}\;,$
where $K_V$ normalises the distribution to unity within the virial 
radius \Rvirh\ of the MW. The total number of sub-halos $N_0$ depends on 
\alpham\ and \Mmin, but there are constraints from current N-body simulations. 
Indeed, there is a rather good agreement about the counting of well resolved 
sub-halos inside a MW-type host: $\sim 100$ objects are found in the mass 
range $10^8-10^{10}\Msol$. Therefore, we ask for 
$N_0 \times\int_{10^8\Msol}^{10^{10}\Msol} d\prob_M/dM = 100$. The inner 
profile \rhocl\ of any sub-halo is also described with 
the NFW or the Moore models, the latter defining a maximally optimised 
scenario. Besides, we need additional prescriptions in order to set the 
scale radius $r_s$ and the associated scale density $\rho_s$ for sub-halos. 
The standard method is to define some reference quantities and to connect them 
with the \emph{physical} scale variables by some functions usually fitted on 
the N-body simulation results. Those reference quantities are the 
\emph{virial} mass and radius, which are related through:
$\Mvir = \frac{4\pi}{3} \Rvir^3 \times 
\left(\Dvir(z=0)\Omega_M(z=0)\rhocrit(z=0)\right)$.
Such a definition is not physical in the sense that any mass is related 
to a radius enclosing \Dvir\ times the background matter density today 
(we set \Dvir=340). Indeed, sub-halos do not evolve in a constant background, 
but inside host halos where the density is not the background density. 
Moreover, 
the previous equation does not carry the formation history of the 
structure, e.g. the fact that lighter clumps should be denser 
(because formed in a denser universe). This is actually encoded in the 
\emph{concentration} parameter $\cvir$, which connects the virial radius, as 
defined above from the mass, to the more physical scale radius:
$\cvir = \frac{\Rvir}{r_{-2}} \;,$
where $r_{-2}$ is the radius at which 
$d/dr \left( r^2 \rhocl(r) \right)|_{r=r_{-2}} = 0$ ($r_{-2} = r_s$ for an NFW 
profile). There are different 
concentration models in the literature, and we have taken two extreme 
cases that encompass them (i) the | maximal | Bullock et 
al~\cite{bullock_etal_01} model (B01) and (ii) the | minimal | Eke et 
al~\cite{eke_etal_01} model (ENS01). Once the sub-halo mass and the 
mass-concentration relation are known, all the clump properties are fixed, 
and one can compute the corresponding annihilation rate. As it is 
proportional to the squared DM sub-halo density, it is useful to define an 
\emph{effective annihilation volume} for any sub-halo as follows:
$\xi \equiv \int_{\rm cl} d^3\vec{x} \left( \frac{\rhocl}{\rhosol}
\right)^2\;,$
where \rhosol, the local dark matter density, allows for a 
normalisation to the local annihilation rate. Such an effective volume 
is that of the DM within the clump would have if it were diluted 
down to the density $\rho_\odot$. It is convenient to normalise to local 
quantities because cosmic ray propagation favours a local origin for the 
exotic contribution. Therefore, $\xi = f(M)$, and 
the mass pdf $d\prob_M/dM$ translates very simply to an annihilation rate pdf. 
Changing the inner profile results in a shift by a constant factor, e.g. 
$\xi_{\rm moore} \simeq 10 \times \xi_{\rm nfw}$, whereas changing the 
concentration model is not as straightforward. Nevertheless, we find very 
roughly that $\xi_{\rm B01} \simeq 1-10 \times \xi_{\rm ENS01}$, depending 
on the sub-halo mass.
\paragraph{Cosmic ray propagation:}
The CR propagation modelling is a key ingredient for this kind of 
studies. Here, we adopt a slab diffusion zone, featured by its radial 
extension that we fix to $R_{\rm slab}=30$ kpc, and by its half-thickness 
$L$, $3$ kpc here. CRs are either confined within the slab or escape forever, 
which is merely fulfilled by imposing Dirichlet boundary conditions to the 
diffusion equation (the CR number density vanishes on the borders). Regarding 
the transport processes, the spatial independent diffusion coefficient is 
given by $K(E)=\beta K_0{\cal R}^{\delta}$ (where ${\cal R}=pc/Ze$ is the 
rigidity) and a constant convective wind $V_{\rm conv}$ is directed outwards 
along the vertical axis. Such a configuration is quantified with the 
\emph{medium} set of parameters provided by~\cite{maurin_etal_01}: 
$K_0 = 0.0112$ kpc$^2$/Myr, $\delta = 0.7$ and $V_{\rm conv} = 12$ km/s. 
One can easily write and solve the diffusion equations for both \poss\ and 
\pbars\ for this kind of geometry. For \poss, the main processes that come 
into play are the energy losses 
(mainly inverse Compton diffusion off CMB or IR photons, and synchrotron 
radiation), and the diffusion on the magnetic turbulences. Disregarding 
the convection process, which is much less efficient than energy losses, 
one can express the typical propagation length for \poss\ as:
$\lambda_{\rm d} \equiv \left(2 K_0 \tau_E 
\left( \frac{\epsilon^{\delta - 1} -
  \epsilon_{S}^{\delta-1}}{1-\delta}\right)\right)^{1/2}\;,$
where $\epsilon\equiv E/\{E_0 = 1\;{\rm GeV}\}$. Assuming an infinite 
3D spherical diffusion zone (correct while $\lambda_{\rm d}\lesssim L$), the 
\pos\ propagator is proportional to a Gaussian function of the source 
distance with $\sigma = \lambda_{\rm d}$ ($\lesssim$ few kpc): Sources located 
farther than $\lambda_d$ will almost not contribute to the flux at the 
Earth. $\lambda_d$ being a decreasing function of the detected 
energy, the effective volume in which \poss\ propagate increases as they 
loose energy. The propagation of \pbars\ must include spallation processes and 
wind convection that occur in the thin Galactic disc, so we can not 
use a simple spherical symmetry to derive a global expression. Moreover, 
the energy losses are negligible for this species, which modifies 
significantly the picture that we had for \poss. Nevertheless, it is 
useful to write, as for \poss, the typical propagation length:
$\Lambda_{\rm d} \equiv \frac{K(E)}{V_{\rm conv}}\;,$
where convection is assumed to dominate over spallation in average, which is 
correct unless at sub-GeV energies. This is quite different from \poss\ 
because this length is an increasing function of energy. 
The picture is therefore reversed, and the propagation volume is much larger 
at higher energy for \pbars\ ($\Lambda_{\rm d}$ reaches 
the size of the diffusion slab at energies of order 10-100 GeV). We now define 
a convenient Green function for any CR species, that 
encodes the injected spectrum induced by dark matter annihilation:
$\gtilde (E,\xsol \leftarrow \vecx) \equiv \int_{E}^{E_{\rm max}} 
dE_S \; {\cal G} (E,\xsol \leftarrow E_S,\vecx)\times \frac{dN_{\rm CR}(E_S)}
{dE_S}\;,$
where $dN_{\rm CR}/dE_S$ is the injected spectrum at source ($E = E_S$ for 
\pbars).
\paragraph{Exotic fluxes and associated boost factors:}
The Galactic host halo is described by a smooth dark matter distribution 
$\rho_{\rm sm}$, so that the corresponding primary CR flux reads:
$\phi_{\rm sm}(E) = \frac{v}{4\pi}{\cal S}\int_{\rm halo} \dvecx \;$ 
$\gtilde (E,\xsol \leftarrow \vecx) \left(\frac{\rho_{\rm sm}}{\rhosol}
\right)^2$
where $v$ is the CR velocity and ${\cal S}\equiv \delta \langle 
\sigma_{\rm ann} v\rangle \rhosol^2/(2\mchi^2)$ encodes the main WIMP 
properties\footnote{$\delta = 1/2$ for Majorana particles, 1 otherwise.}.
Sub-halos can be considered as point-like sources, and the flux due to 
the $i^{\rm th}$ object is merely:
$\phi_{{\rm cl},i} (E) = \frac{v}{4\pi}\times {\cal S} \times \xi_i \times 
\gtilde (E,\xsol \leftarrow \vecx_i)$.
Then, we have to sum over the whole population. We can derive a 
statistical prediction by integrating over the sub-halo phase space.
The overall clump contribution is thus:
$\phi_{\rm cl,tot} (E) =$ $\frac{v}{4\pi} N_0  
\int dM \xi(M) \frac{d{\cal P}_M(M)}{dM}\times $ $\int\dvecx 
\gtilde(E,\xsol \leftarrow \vecx) \frac{d{\cal P}_V(\vecx)}{\dvecx}
 =$  $ N_0 \langle \phi_{\rm cl} \rangle = 
\frac{v}{4\pi}{\cal S}  N_0  \langle \xi \rangle 
\langle \gtilde \rangle$
where the 1$^{\rm st}$ equality takes directly the | normalised | 
pdfs into account (this limit would be reached e.g. for an infinite 
number of MC realisations). Those pdfs characterise 
$\xi$ and \gtilde, the former being a function of the sub-halo mass and the 
latter being spatially weighted with $d{\cal P}_V/dV$. The last equality gives 
the same quantities in terms of statistical mean values. This assumes no 
correlations between the considered variables. The calculation of the 
variance of CR fluxes originating from sub-halos 
$\sigma_{\rm cl,tot}$ is straightforward:
$\frac{\sigma_{\rm cl,tot}^2}{\phi_{\rm cl,tot}^2} = 
\frac{1}{N_0 } \left(\frac{\sigma_\xi^2}{\langle\xi\rangle^2}+ 
\frac{\sigma_\gtilde^2}{\langle\gtilde\rangle^2} +
\frac{\sigma_\xi^2\sigma_\gtilde^2}{\langle\xi\rangle^2\langle\gtilde\rangle^2}
\right) \;,$
where $\sigma_x$ is the variance corresponding to any variable $x$. The 
\emph{boost factors} for \poss\ and \pbars\ are the ratios 
of the CR fluxes originating from a clumpy halo to those 
calculated for the host smooth halo alone. Normalising the whole dark matter 
average density profile at the Earth, the effective boost is:
$\Beff(E)= (1-\fsol)^2 + \phi_{\rm cl,tot}/\phi_{\rm sm}\;, $
which depends on energy. \fsol\ is the average local DM density in form of 
clumps (the smooth component density is $(1-\fsol)\rhosm$ when sub-halos 
are added). Estimates of the mean fluxes for \poss\ and \pbars\ and associated 
statistical variances have been performed by using the semi-analytical method 
proposed by \cite{boost_method_lavalle_etal_07}. In order to exhaustively 
characterise the effect of all the relevant variables in the problem, we have 
varied the minimal mass of sub-halos \Mmin, the logarithmic slope of the mass 
function \alpham, the concentration model, the inner profiles of sub-halos, 
their spatial distribution in the MW ; we have also varied the CR propagation 
model. This provides a way to anticipate predictions for any clumpiness 
modelling, and is also useful to sketch some theoretical uncertainty contours.
The full results are available in the original 
paper~\cite{boost_clumps_lavalle_etal_07}. In this proceeding, we quickly 
discuss three cases: \emph{minimal}, \emph{reference} and \emph{maximal} 
models, considering a medium set of CR propagation parameters. By 
defining a model with $\{\Mmin,\alpham,$ inner profile, concentration model, 
spatial distribution$\}$, the \emph{minimal} 
reads $\{10^6\Msol$,1.8,NFW,ENS01,cored $\}$, the 
\emph{reference} $\{10^{-6}\Msol$,1.9,NFW,B01,cored $\}$ 
and the \emph{maximal} 
$\{10^{-6}\Msol$,2.0,Moore,B01,smooth-tracking $\}$. The most 
optimistic | \emph{maximal} | case, which is also the most unlikely, yields 
boost factors $\sim 20$, which depend on the CR energy, while the other 
configurations do not permit substantial enhancement, the effective boost 
factors \Beff\ remaining close to unity for both \poss\ and 
\pbars. Taking back NFW profiles in the 
\emph{maximal} setup, we would obtain $\sim 20/10 = 2$. The variance affecting 
those predictions decreases with the local average number density of 
sub-halos. Indeed, the statistical variance depends on the number of objects 
located inside a volume bounded by the CR propagation length and contributing 
to the flux at the Earth. As the propagation length is a function of the CR 
energy which depends on the species | increasing (respectively decreasing) 
with energy for \pbars\ (\poss) | one can understand why the picture 
can be different for different CR species. As a very general statement, the 
variance on primary fluxes is smaller at lower energy for \poss, or at 
higher energy for \pbars. Nevertheless, even with a large variance for 
sub-halo fluxes, the statistical error will still be small for boost factors 
as soon as the smooth contribution will dominate (i.e.~errors are small for 
small boost factors).

\paragraph{Conclusion:}
In~\cite{boost_clumps_lavalle_etal_07}, we have tried to exhaustively tackle 
the problem of DM clumpiness effects in the frame of indirect detection of 
annihilating DM with antimatter cosmic rays. We have taken the whole phase 
space of sub-halos into account, using a compilation of results found in the 
literature, coming from N-body simulations as well as from analytical models. 
We have sketched the theoretical uncertainties affecting the useful dark 
matter parameters, and defined rather wide associated ranges to account for 
them. For the sake of completeness, we have also considered different sets of 
CR propagation parameters, which are still degenerate and not enough 
constrained by the existing secondary/primary CR measurements. 
We refer the reader to the original paper for very detailed results. In 
this proceeding, we have only illustrated the cases of extreme configurations, 
which can provide boost factors $\lesssim 20$ for both \poss\ and 
\pbars. Nevertheless, our analysis strongly disfavours large and even 
mildly boost factors for most parts of the large parameter space used to 
describe sub-halos. This is almost independent of CR propagation uncertainties.

\paragraph{Acknowledgements:} The author is grateful to his collaborators 
X.-J.~Bi, D.~Maurin and Q.~Yuan, for interesting discussions during this 
exciting work. It is also a pleasure to thank P.~Salati and R.~Taillet for 
early fruitful collaborations on the topic, and for motivating on-going 
exchanges.

\end{document}